\documentclass{article}                    
\usepackage{graphicx}
\usepackage[square,numbers]{natbib}
\usepackage{hyperref}
\newcommand{\ra}{\rangle}
\newcommand{\la}{\langle}
\begin{document}

\title{Spin Path Integrals and Generations}
\author{Carl Brannen\\
              8500 148th Ave. NE, T-1064, \\
              Redmond, WA 98052 USA\\
              carl@brannenworks.com}

\maketitle

\begin{abstract}
The spin of a free electron is stable but its position is not. Recent quantum information research by G. Svetlichny, J. Tolar, and G. Chadzitaskos have shown that the Feynman \emph{position} path integral can be mathematically defined as a product of incompatible states; that is, as a product of mutually unbiased bases (MUBs). Since the more common use of MUBs is in finite dimensional Hilbert spaces, this raises the question ``what happens when \emph{spin} path integrals are computed over products of MUBs?'' Such an assumption makes spin no longer stable. We show that the usual spin-1/2 is obtained in the long-time limit in three orthogonal solutions that we associate with the three elementary particle generations. We give applications to the masses of the elementary leptons.

\end{abstract}

$\;$ 

The first section discusses mutually unbiased bases and position path integrals, and the difference in behavior between position and spin. Section 2 introduces spin path integrals over MUBs. Section 3 derives some convenient arithmetic results for products of spin-1/2 projection operators. Section 4 calculates the long-time MUB spin path integrals. Section 5 shows that the long time propagators, after summing over orientation, converge to the usual spin-1/2. Section 6 applies the results to the lepton masses. Finally, section 7 discusses the results.

\section{Introduction}\label{sec:intro}

Let $A = \{|a_j\ra\}$ and $B = \{|b_k\ra\}$ be two different bases for a finite dimensional Hilbert space. They are ``mutually unbiased'' if all inner products $\la a_j|b_k\ra$, have the same magnitude. If the dimension of the Hilbert space is $N$, then the inner products have magnitude $\sqrt{1/N}$:
\begin{equation}
|\la a_j | b_k\ra| = \sqrt{1/N}.
\end{equation}
The simplest example of mutually unbiased bases are spin-1/2 in two perpendicular directions. Bases for spin-1/2 in the $\vec{x}, \vec{y}$ and $\vec{z}$ direction are:
\begin{equation}
\begin{array}{ccc}\label{eq:MUBbasis}
\left\{\;|+\vec{x}\ra,\; |-\vec{x}\ra\;\right\} &=& \left\{
\left(\begin{array}{c}\sqrt{1/2}\\ \sqrt{1/2}\end{array}\right),
\left(\begin{array}{c}\sqrt{1/2}\\-\sqrt{1/2}\end{array}\right)\right\},\\
\left\{\;|+\vec{y}\ra,\; |-\vec{y}\ra\;\right\} &=& \left\{
\left(\begin{array}{c}\sqrt{1/2}\\ i\sqrt{1/2}\end{array}\right),
\left(\begin{array}{c}\sqrt{1/2}\\-i\sqrt{1/2}\end{array}\right)\right\},\\
\left\{\;|+\vec{z}\ra,\; |-\vec{z}\ra\;\right\} &=& \left\{
\left(\begin{array}{c}1\\0\end{array}\right),
\left(\begin{array}{c}0\\1\end{array}\right)\right\}.
\end{array}
\end{equation}
These three bases are mutually unbiased; the magnitudes of the transition amplitudes are all $\sqrt{1/N} = \sqrt{1/2}$. This is a complete set; there are only three perpendicular directions in three dimensions.

Zee's textbook introduction to quantum field theory \cite{QFTinaNutShell} introduces the path integral formulation with the double slit experiment. ``A particle emitted from a source $S$ at time $t=0$ passes through one or the other of two holes, $A_1$ and $A_2$ drilled in a screen and is detected at time $t=T$ by a detector located at $O$. The amplitude for detection is given by a fundamental postulate of quantum mechanics, the superposition principle, as the sum of the amplitude for the particle to propagate from the source $S$ through the hole $A_1$ and then onward to the point $O$ and the amplitude for the particle to propagate from the source $S$ through the hole $A_2$ and then onward to the point $O$.''

\begin{figure}[!htp]
\setlength{\unitlength}{1pt} %
\begin{picture}(220,100)
\thinlines
\put(25, 5){\vector(0,1){90}}
\put(22,30){\line(1,0){6}}
\put(22,50){\line(1,0){6}}
\put(22,70){\line(1,0){6}}
\put( 8,30){1}
\put( 8,50){2}
\put( 8,70){3}
\put(28,90){position}
\put( 48,60){$S$}
\put( 50,50){\circle*{5}}
\put(208,60){$O$}
\put(210,50){\circle*{5}}
\thicklines %
\multiput(90,20)(40,0){3}{\line( 0, 1){8}}
\multiput(90,32)(40,0){3}{\line( 0, 1){16}}
\multiput(90,52)(40,0){3}{\line( 0, 1){16}}
\multiput(90,72)(40,0){3}{\line( 0, 1){8}}
\put( 90, 82){$A$}
\put(130, 82){$B$}
\put(170, 82){$C$}
\thinlines
\multiput(50,50)(40,0){3}{\qbezier[24](0,0)(20,-10)(40,-20)}
\multiput(50,50)(40,0){3}{\qbezier[24](0,0)(20, 10)(40, 20)}
\multiput(50,50)(40,0){4}{\qbezier[20](0,0)(20,  0)(40,  0)}
\multiput(90,30)(40,0){2}{\qbezier[20](0,0)(20,  0)(40,  0)}
\multiput(90,70)(40,0){2}{\qbezier[20](0,0)(20,  0)(40,  0)}
\multiput(90,30)(40,0){2}{\qbezier[28](0,0)(20, 20)(40, 40)}
\multiput(90,70)(40,0){2}{\qbezier[28](0,0)(20,-20)(40,-40)}
\multiput(90,30)(40,0){3}{\qbezier[24](0,0)(20, 10)(40, 20)}
\multiput(90,70)(40,0){3}{\qbezier[24](0,0)(20,-10)(40,-20)}
\end{picture}
\caption{\label{fig:3x3paths} Paths from $S$ to $O$ through three intermediate screens $A$, $B$, and $C$, with three holes each.}
\end{figure}
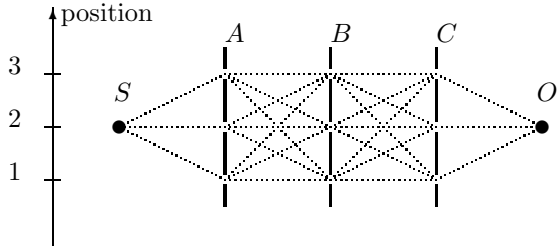

Increasing the number of holes increases the number of paths. Thus with three holes the total amplitude is the sum of three single path amplitudes:
\begin{equation}
\mathcal{A}(S\to O) = \Sigma_{j=1}^3 \mathcal{A}(S\to A_j \to O).
\end{equation}
Adding two more screens $B$ and $C$, between $A$ and $O$, see Fig.~(\ref{fig:3x3paths}), requires summing the amplitudes over $3^3 =  27$ paths:
\begin{equation}
\mathcal{A}(S\to O) = \Sigma_{j=1}^3\Sigma_{k=1}^3\Sigma_{l=1}^3 \mathcal{A}(S\to A_j \to B_k \to C_l \to O).
\end{equation}
The path integral formalism follows by considering increases in the number of intermediate screens and the number of holes drilled in them. If the screens have enough holes, and are sufficiently closely spaced, one requires that the calculation should give the transition amplitude for a free particle moving in empty space between $S$ and $O$.

For a particle moving in a potential $V(\hat{q})$, the final result for the amplitude is
\begin{equation}
\la q_T|e^{-iHT}|q_0\ra = \int Dq(t)e^{i\int_0^T dt [\frac{1}{2}m\dot{q}^2 - V(q)]}.
\end{equation}
In the above $Dq(t)$ represents all possible paths $q$ which run from the source $S = q(0)$ to the detector $O = q(T)$. The contribution from each path consists of a complex phase.

In 2008, G. Svetlichny \cite{Svet2008} noticed that while MUBs and path integrals appear to be very different things, they both involve amplitudes where the information is contained only in the phases. He showed that in the limit of short time intervals, the path integral approaches the transition amplitudes between two MUBs. Thus longer paths consist of products of MUB transition amplitudes. In 2009, J. Tolar and G. Chadzitaskos \cite{Tolar2009} obtained the free particle propagator as a sum over products of MUB transition amplitudes by a limiting procedure in finite dimensional Hilbert spaces.

To see quantum behavior in the motion of an electron we must measure its position with an accuracy smaller than its de Broglie wavelength: \cite{Messiah}
\begin{equation}
\lambda = \frac{12.2}{\sqrt{E_{(\textrm{eV})}}} \; \textrm{\AA}.
\end{equation}
For an electron with energy 1 keV, this distance is $\lambda = 0.4 \; \textrm{\AA} = 4\times 10^{-11}m$. Of the particle detectors we have available, the most accurate is emulsion which can measure particle positions to an accuracy of around $5\times 10^{-7}m$. \cite{RoePP} This is 4 orders of magnitude larger than the de Broglie wavelength of a 1 keV electron. Heavier and higher energy particles have even smaller de Broglie wavelengths. Consequently, elementary particle tracks appear classical. \cite{LandLQM} Instead, the best evidence we have for the bizarre behavior of quantum particles over short times and accurate positions is obtained from diffraction experiments such as the single slit and double slit experiments.

In the very short time limit, a product of MUB transition amplitudes approaches a single MUB transition amplitude. In this case all possible transitions are equally probable. This behavior corresponds to the familiar result of single slit experiments: with a sufficiently narrow slit, the particles receive random velocities and their tracks spread out.

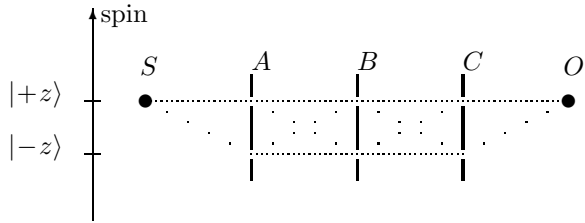
\begin{figure}[!htp]
\setlength{\unitlength}{1pt} %
\begin{picture}(250,90)
\thinlines
\put(40, 5){\vector(0,1){80}}
\put(37,30){\line(1,0){6}}
\put(37,50){\line(1,0){6}}
\put( 8,30){$|\!-\!z\ra$}
\put( 8,50){$|\!+\!z\ra$}
\put(43,80){spin}
\put( 58,60){$S$}
\put( 60,50){\circle*{5}}
\put(218,60){$O$}
\put(220,50){\circle*{5}}
\thicklines %
\multiput(100,20)(40,0){3}{\line( 0, 1){8}}
\multiput(100,32)(40,0){3}{\line( 0, 1){16}}
\multiput(100,52)(40,0){3}{\line( 0, 1){8}}
\put(100, 62){$A$}
\put(140, 62){$B$}
\put(180, 62){$C$}
\thinlines
\multiput( 60,50)(40,0){3}{\qbezier[ 5](0,0)(20,-10)(40,-20)}
\multiput( 60,50)(40,0){4}{\qbezier[20](0,0)(20,  0)(40,  0)}
\multiput(100,30)(40,0){2}{\qbezier[20](0,0)(20,  0)(40,  0)}
\multiput(100,30)(40,0){3}{\qbezier[ 5](0,0)(20, 10)(40, 20)}
\end{picture}
\caption{\label{fig:NoCrosses} A free particle beginning with spin-up stays that way; forbidden transitions shown with sparse dots.}
\end{figure}

The free propagator for a spin-1/2 particle does not change spin; that is, the transition amplitudes between spin-up $|+\vec{z}\ra$, and spin-down $|-\vec{z}\ra$ are zero. In terms of path integrals over spin space, the paths do not cross, spin-up stays spin-up. See Fig.~(\ref{fig:NoCrosses}). If we measure the spin of a beam of spin-up electrons the result is always spin-up. This paper considers the possibility that if we could measure spin over a sufficiently short time interval, we would find the same behavior as position: the measurement of spin would modify the spin. Thus the traditional Stern-Gerlach measurement of spin-1/2 is classical in the sense that continuous particle tracks are classical. We assume that the underlying quantum behavior consists of transitions between mutually unbiased bases. For an interesting argument that the Stern-Gerlach experiments can be interpreted entirely from a classical understanding of electricity and magnetism, see \cite{Franca98,Franca09}.

The effort here is similar (and uses similar mathematics) to that of Foster and Jacobson; see \cite{FosterAndJacobson} and references therein. They attempt to obtain the massive Dirac propagator from the chiral left and right handed states. In doing this, they consider paths that move along one of four directions in spacetime. The steps are oriented towards the corners of a tetrahedron in 3 dimensions and move one unit forwards in time. Since they are replacing the usual two-element basis set for spin-1/2 with a set of size four, one would expect them to obtain two copies of the Dirac propagator. They avoid the unwanted doubling by making computations that begin and end with spinors. Thus their propagators amount to $2\times 2$ matrices and describe only the desired two states. These ideas originated with Feynman's ``checkerboard'' model of the electron in $1+1$ dimensions. \cite{Checkerboard}.

The present paper's calculation begins with a tripled basis set for spin-1/2 and since we're looking for three generations of massless chiral fermions, we do not avoid the tripling. Instead of spinors, our calculations concentrate on what happens to the propagators at long time. In addition, we work in the quantum information approximation; we do not consider position.

\section{Spin Projection Operators}\label{sec:SpinProjOperators}

For the Hilbert space with dimension M=2, Svetlichny's prescription for the analog to the Feynman path integral is (see equation (9) of \cite{Svet2008}):
\begin{equation}\label{eq:SvetNumberNine}
\la q_T|q_0\ra = \Sigma_{x_1}\Sigma_{x_2}...\Sigma_{x_N}
\la q_T|x_1\ra\la x_1|x_2\ra ... \la x_{N-1}|x_N\ra \la x_N|q_0\ra
\end{equation}
where we require that $x_j$ and $x_{j+1}$ be mutually unbiased for all $j$. This sum over products reduces since for each $j$, we have that the set $\{x_j\}$ are a complete set: $\Sigma_{x_j}|x_j\ra\la x_j| = 1$.

Equation (\ref{eq:SvetNumberNine}) is unsatisfactory in that it implies a fixed sequence of choices of basis state. For example, the first sum $\Sigma_{x_1}$, might be over spin in the $\pm x$ direction, the next sum over spin in the $\pm z$, etc. While it may be mathematically true, it is not physically satisfying.

If we eliminated the arbitrary choice of basis by allowing $x_j$ to run over any of the six states in the three MUBs given in Eq.~(\ref{eq:MUBbasis}) we would eliminate the arbitrary choice of bases, but the calculation would still give a trivial result since $\Sigma_{x_j} |x_j\ra\la x_j| = 3$. Instead we will require that $x_j$ be taken from one of the three positive spin states:
\begin{equation}
|+x\ra,\;\;\;|+y\ra,\;\;\;|+z\ra.
\end{equation}
Thus instead of summing over the two states taken from one of the three (arbitrarily chosen) basis states, we will allow our sums to run over three basis states, one chosen from each basis.

We will be summing quantities such as the following:
\begin{equation}\label{eq:ExMUBtranAmp}
\la +\vec{x}|+\vec{z}\ra \la +\vec{z}|+\vec{x}\ra \la +\vec{x}|+\vec{y}\ra \la +\vec{y}|+\vec{z}\ra.
\end{equation}
Starting on the right, this corresponds to a particle whose spin path goes through the sequence $|+\vec{z}\ra$ to $|+\vec{y}\ra$, to $|+\vec{x}\ra$, to $|+\vec{z}\ra$, and finally to $|+\vec{x}\ra$. To make these calculations, we will rewrite the above as:
\begin{equation}
\la +\vec{x}|\;\;|+\vec{z}\ra \la +\vec{z}|\;\;|+\vec{x}\ra \la +\vec{x}|\;\;|+\vec{y}\ra \la +\vec{y}|\;|+\vec{z}\ra.
\end{equation}

The operator $|+\vec{x}\ra\la+\vec{x}|$ is the projection operator for spin in the $+\vec{x}$ direction and similarly for $+\vec{y}$ and $+\vec{z}$. We will label these spin projection operators $X$, $Y$, and $Z$. In terms of the Pauli spin matrices, they can be written as:
\begin{equation}
\begin{array}{rclcl}
X &=& |+\vec{x}\ra \la +\vec{x}| &=& (1 + \sigma_x)/2,\\
Y &=& |+\vec{y}\ra \la +\vec{y}| &=& (1 + \sigma_y)/2,\\
Z &=& |+\vec{z}\ra \la +\vec{z}| &=& (1 + \sigma_z)/2.
\end{array}
\end{equation}
In this notation, Eq.~(\ref{eq:ExMUBtranAmp}) becomes $\la +\vec{x}| XZXYZ |+\vec{z}\ra$.

More generally, for any vector $\vec{u}$, define $\sigma_u = u_x\sigma_x + u_y\sigma_y + u_z\sigma_z$. Then $(1+\sigma_u)/2$ is the projection operator for spin-1/2 in the $+\vec{u}$ direction, and $(1-\sigma_u)/2$ projects spin-1/2 in the $-\vec{u}$ direction. We will use a bar for the opposite spin state, $\bar{U} = (1-\sigma_u)/2$.

When one converts a basis set $\{|+\vec{u}\ra, |-\vec{u}\ra\}$ into projection operators, the features which characterize a basis set are transformed into algebraic relations among the projection operators. The normality of the kets become idempotency:
\begin{equation}\label{eq:Idempotency}
\langle \pm \vec{u}|\pm \vec{u}\rangle = 1 \to [(1\pm \sigma_u)/2]\;[(1\pm \sigma_u)/2] = (1\pm \sigma_u)/2,
\end{equation}
and orthogonality becomes annihilation:
\begin{equation}\label{eq:Annihilation}
\langle \pm\vec{u}| \mp\vec{u}\rangle = 0 \to [(1 \pm \sigma_u)/2]\;[(1 \mp \sigma_u)/2] = 0.
\end{equation}
Finally, the requirement that the number of basis elements equals the dimensionality of the vector space becomes the requirement that the sum of the projection operators is unity:
\begin{equation}\label{eq:SumsTo1}
(1 +\sigma_u)/2 + (1-\sigma_u)/2 = 1.
\end{equation}
In addition the projection operators have trace 1:
\begin{equation}\label{eq:TraceIs1}
\textrm{tr}((1\pm\sigma_u)/2) = \textrm{tr}(1/2) \pm \textrm{tr}(\sigma_u/2) = 1.
\end{equation}
We will find the same relations among the long-time MUB spin path integrals. Our set of annihilating projection operators that sum to unity will have three basis elements; we will associate these with the three generations of elementary spin-1/2 fermions.

\section{Spin Projection Arithmetic}\label{sec:MubArith}

There are three products of three spin projection operators that begin and end with $Z$:
\begin{equation}
\begin{array}{rcr}
Z\;X\;Z &=& \sqrt{1/2^2}Z,\\
Z\;Y\;Z &=& \sqrt{1/2^2}Z,\\
Z\;Z\;Z &=& \sqrt{1/2^0}Z.
\end{array}
\end{equation}
These are real multiples of $Z$. The powers of two in the above equations count the number of transitions. With $ZXZ$ and $ZYZ$ there are two transitions, and with $Z Z Z = Z$ there are none. More general products can be complex:
\begin{equation}\label{eq:ZXYZ}
Z\;X\;Y\;Z = \sqrt{+i/2^3}\;Z,
\end{equation}
where we define $\sqrt{\pm i} = \exp(\pm i\pi/4)$. A path that makes $N$ transitions through different projection operators will have a magnitude of $\sqrt{2^{-N}}$.

The complex phase $\sqrt{+i}$ in Eq.~(\ref{eq:ZXYZ}) is a geometric phase, also called Berry \cite{Berry1984} or Pancharatnam \cite{Pancharatnam} phase. These phases can be picked up when a quantum particle goes through a series of states and returns to its initial state. It does not depend on the arbitrary complex phases of spinors and consequently is an observable. Our use of quantum phase will arise from products of projection operators of the sort described in R. Bhandari's paper \cite{Bhandri1998}, but for spin-1/2 rather than photon polarization.

In computing sums of MUB Feynman paths, we will use $X$, $Y$, and $Z$ as part of a basis for a complex vector space. These three account for paths that begin and end with the same projection operator. Paths whose final projection operator are different from their initial projection operator must be handled differently. For these mixed paths there are six ways to choose different initial and final projection operators.

There are three possible products of three projection operators that begin with $Z$ (on the right) and end with $X$:
\begin{equation}
\begin{array}{rcr}
X\;X\;Z &=& \sqrt{ 1/2^0}\;XZ,\\[0 pt]
X\;Y\;Z &=& \sqrt{+i/2^1}\;XZ,\\[0 pt]
X\;Z\;Z &=& \sqrt{ 1/2^0}\;XZ.
\end{array}
\end{equation}
In the above calculation, the three products are all complex multiples of the same matrix: $XZ$. Twice this matrix is idempotent and has trace 1:
\begin{equation}\label{eq:XZprod}
(2XZ)^2 = 2XZ = \left(\begin{array}{cc}
1&0\\
1&0\end{array}\right).
\end{equation}
This is a property of all MUB paths that begin and end with different projection operators; they will be complex multiples of an idempotent with unit trace.

Products that begin and end with orthogonal projection operators are the familiar raising and lowering operators. For example $ZX\bar{Z}$ and $ZY\bar{Z}$ are both raising operators for spin-1/2. They differ by a complex phase:
\begin{equation}
ZX\bar{Z} = e^{i\pi/2}ZY\bar{Z}.
\end{equation}
The path $ZX\bar{Z}\;\bar{Z}YZ$ picks up the same complex phase, $\pi/2$:
\begin{equation}
ZX\bar{Z}YZ = e^{i\pi/2}\sqrt{1/2^4}\;Z.
\end{equation}
Thus the choice of phase in raising and lowering operators are an example of geometric phase.

The pure density matrices are usually defined as the matrices that can be produced from normalized spinors. An alternative definition is that they are the Hermitian projection operators with trace 1. To convert a pure density matrix to a spinor choose a nonzero column (or the complex conjugate of a nonzero row), treat the elements as a vector, and normalize. The matrix used in Eq.~(\ref{eq:XZprod}), and the other five obtained by replacing $X$ and $Z$ with different projection operators, possesses two of the three properties that define a pure density matrix; they are idempotent and have trace 1 but they are not Hermitian; their rows give different spinors than their columns.

Hermiticity is associated with the property of time reversal invariance. We have:
\begin{equation}
(2XZ)^{\dag} = 2ZX.
\end{equation}
More generally, Hermitian conjugation reverses the order of a product of any number of projection operators. This corresponds to the reversal of the time ordering of a path.

For convenience, we will drop off the factor of 2 and use 
\begin{equation}\label{eq:PathBasis}
\begin{array}{ccccccccc}
\{X,&Y,&Z,&XY,&YX,&XZ,&ZX,&YZ,&ZY\}
\end{array}
\end{equation}
as the basis for MUB spin path integrals. Any product of MUB projection operators can be written as a complex multiple of one of these nine. In using these, we have five more than are needed to give a basis for $2\times 2$ complex matrices, but what we are looking for is a basis for paths and path integral calculations.

In Feynman path integrals, one specifies initial and final states of the particles and sums over diagrams that connect those states. For the MUB case we will do the same. For this reason, our set of nine products of projection operators can be used as if they were a basis set for a complex vector space of dimension nine. A vector in that space is a collection of nine amplitudes. Adding two such vectors together is equivalent to adding nine Feynman path integrals to nine other Feynman path integrals to get nine sums. In this way, given a path, (or a collection of MUB paths that all share the same initial and final projection operators), we associate a complex number.

In addition to summing two paths, we also need to connect one path to another. This is equivalent to the multiplication of two products of projection operators, for example, $(XY)(YZ)$ $=(XYYZ)$. Suppose we have two paths (or sums of paths) $A$ and $B$ which are associated with complex numbers $a$ and $b$. If it so happens that path $A$ ends with the same projection operator that path $B$ begins with, then the concatenation of these paths, call it $BA$, will be associated with the complex number $ba = ab$ multiplied by a non commutative correction. The correction depends only on the initial and final projection operators of the paths. This suggests that we should organize our calculations so that transitions between projection operators are kept inside products of the path integral basis Eq.~(\ref{eq:PathBasis}); we can then use complex multiplication to model the concatenation of paths.

Accordingly, we will only concatenate two path integral basis elements if the final projection operator of the first element matches the initial projection operator of the second element. For example, this requirement allows $(XY)(YZ)$ or $(Z)(ZX)$ but not $(XY)(Z)$ or $(Z)(X)$. Since the path integral basis has nine elements, there are a total of 81 possible products of them but the restriction reduces the number of products we will consider to 27. Below, we will show that this is sufficient to account for any spin path, see Eq.~(\ref{eq:MakeLegal}).

Of the 27 remaining products of path integral basis elements, 15 are already correct for standard complex multiplication. The five that begin with $X$ are:
\begin{equation}
\begin{array}{rcl}
(X)(X) &=& X,\\
(X)(XY) &=& XY,\\
(X)(XZ) &=& XZ,\\
(XY)(Y) &=& XY,\\
(XZ)(Z) &=& XZ.
\end{array}
\end{equation}
The other ten such products are obtained by cyclic permutation of $X$, $Y$, and $Z$. We will call these the ``diagonal products'' for reasons which will be clear below. The remaining 12, ``off diagonal products'' are more complicated. The four that begin with $X$ are:
\begin{equation}\label{eq:offdiagonal}
\begin{array}{rcl}
(XY)(YX) &=& \sqrt{ 1/2^2}\; X,\\
(XZ)(ZX) &=& \sqrt{ 1/2^2}\; X,\\
(XY)(YZ) &=& \sqrt{+i/2^1}\; XZ,\\
(XZ)(ZY) &=& \sqrt{-i/2^1}\; XY.
\end{array}
\end{equation}
The other eight are obtained by cyclic permutation.

Other than the complex factors of the off diagonal products, the above 27 products are compatible with matrix multiplication. Let $(a'_x,a'_y,...a'_{yz})$ be a 9-vector of complex numbers associated with a collection of MUB spin paths. We use the prime to indicate that these represent complex multiples of non commutative matrices. For the moment we will suppose that there is some single path associated, perhaps of type $XZ$, so that only one of them (i.e. $a'_{xz}$) is nonzero. Assemble them into a $3\times 3$ matrix $a'$ as follows:
\begin{equation}
a' = \left(\begin{array}{ccc}
a'_x&a'_{xy}&a'_{xz}\\
a'_{yx}&a'_y&a'_{yz}\\
a'_{zx}&a'_{zy}&a'_z
\end{array}\right)
\end{equation}
In this matrix, the columns represent the three initial states $\la+\vec{x}|$, $\la+\vec{y}|$, $\la+\vec{z}|$, while the three rows represent the final states $|+\vec{x}\ra$, $|+\vec{y}\ra$, and $|+\vec{z}\ra$.

Let $b'$ be a similar matrix for another path whose final projection operator matches the initial projection operator of $a'$, that is, $Z$. For instance, $b'$ could represent a path of type $ZY$ so that only $b'_{zy}$ is nonzero. Concatenating the paths gives a path of type $XY$. Corresponding to this, the matrix product $(ab)'$ will have only one nonzero entry, $(ab)'_{xy} = a'_{xz}b'_{zy}$. This is not quite correct; according to Eq.~(\ref{eq:offdiagonal}) we should have $(ab)'_{xy} = \sqrt{-i/2^1}\;a'_{xz}b'_{zy}$.

To fix the matrix product, we need to scale the off diagonal elements (i.e. $a'_{xy}$, $a'_{yz}$, $a'_{zx}$, $a'_{yx}$, $a'_{zy}$, $a'_{xz}$ and similarly for $b'$) in such a way that the off diagonal products are corrected without changing the diagonal products. Rewriting Eq.~(\ref{eq:offdiagonal}) in terms of what it says about products of the elements of $a'$ and $b'$, what we want are new matrices, $a$, $b$, and $(ab)$ such that:
\begin{equation}\label{eq:offa}
\begin{array}{rcl}
\sqrt{ 1/2^2} \; (ab)'_{x}  &=& a'_{xy}b'_{yx},\\
\sqrt{ 1/2^2} \; (ab)'_{x}  &=& a'_{xz}b'_{zx},\\
\sqrt{+i/2^1} \; (ab)'_{xz} &=& a'_{xy}b'_{yz},\\
\sqrt{-i/2^1} \; (ab)'_{xy} &=& a'_{xz}b'_{zy},
\end{array}
\end{equation}
becomes $(ab)_x = a_{xy}b_{yx}$, etc. A suitable transformation is $a' \to a$ by:
\begin{equation}
\begin{array}{rcr}
a_x    &=& a'_x,\\
a_{xy} &=& \eta_g \; a'_{xy},\\
a_{xz} &=& \eta_g^*\; a'_{xz},
\end{array}
\end{equation}
where 
\begin{equation}
\eta_g = \sqrt{1/2}\;e^{+i\pi/12}e^{2ig\pi/3}, \;\;\textrm{for $g=1,2,3$}.
\end{equation}
Cyclic permutations give the transformations on the other six path basis elements. The complex phase $2i\pi/3$ appears repeatedly; we will abbreviate it as $w$:
\begin{equation}
w = \exp(2i\pi/3),
\end{equation}
so $\eta_g = \sqrt{1/2}\;\exp(i\pi/12)\;w^g$. We will use the integer parameter $g$ to represent the generation quantum number.

The matrix entries of $b'$ represent the nine cases: spin changing from $+x$ to $+x$, $+x$ to $+y$, etc. Suppose this propagator is followed by another propagator $a'$. In computing the propagator $(ab)'$ we must sum over all possible paths. With matrices $a$ and $b$, this simply amounts to matrix multiplication. For example, the three terms on the right side of
\begin{equation}
(ab)_{xz} = a_{x}b_{xz} +  a_{xy}b_{yz} + a_{xz}b_{z}.
\end{equation}
correspond to the three paths that go through the new node with spin $+\vec{x}$, $+\vec{y}$, and $+\vec{z}$, respectively. Thus we have transformed the problem of concatenating the propagators of MUB spin into $3\times 3$ matrix multiplication.

\section{Long-Time MUB Spin Propagator}\label{sec:TriplePauliSpinProp}

Let's begin by computing the spin path integral (or propagator) from $+\vec{x}$ to $+\vec{z}$ with two internal paths, call it $G_2(+\vec{x},+\vec{z})$. This is a sum over paths with four projection operators. The initial projection operator is $X$, the final projection operator $Z$ while the two inner projection operators can be any of $X$, $Y$, or $Z$. Taking all possible cases for the inner projection operators and calculating with the Pauli spin matrices, or using the rules given above, we find:
\begin{equation}\label{eq:SumOverPathsExample}
\begin{array}{rcl}
G_2(+\vec{x},+\vec{z}) &=& ZXXX + ZXYX + ZXZX\\
&+&ZYXX + ZYYX + ZYZX\\
&+&ZZXX + ZZYX + ZZZX,\\
&=&(3-3i/4)\left(\begin{array}{cc}
1&1\\0&0\end{array}\right),\\
&=&(6-3i/2)ZX.
\end{array}
\end{equation}
This amplitude is too large to conserve probability.

The problem is that the transition amplitudes need to be adjusted for the fact that we have added two new possible paths at each vertex. To preserve probability at each vertex, we need to make the transition probabilities smaller. There are three transitions so we will multiply by a factor $\kappa^3$.

We will compute $\kappa$ later in this section, for now, let's see how to rewrite a path as a matrix multiplication. We are concerned with paths that look like $(Z)(P)(Q)(X)$ where $(P)$ and $(Q)$ can be any of the three projection operators. First, using idempotency, we duplicate $P$ and $Q$ and factor into pairs:
\begin{equation}\label{eq:MakeLegal}
\begin{array}{rcl}
ZPQX &=& Z(PP)(QQ)X,\\
&=&(ZP)(PQ)(QX).
\end{array}
\end{equation}
Modify the pairs to replace $(XX)=(X)$, $(YY) = (Y)$, and $(ZZ) = (Z)$. Now we have the path as a product of our nine path basis elements in such a way that each pair of adjacent basis elements match their adjacent projection operators.

The state $ZP$ has final state $|+\vec{z}\ra$, and initial state anything, so we represent it as $\kappa$ times the matrix with 1s in the bottom $(z)$ row:
\begin{equation}
(zp)' = \kappa \left(\begin{array}{ccc}
0&0&0\\
0&0&0\\
1&1&1
\end{array}\right).
\end{equation}
Convert the matrix $(zp)'$ to $(zp)$ to obtain:
\begin{equation}
(zp) = \kappa \left(\begin{array}{ccc}
0&0&0\\
0&0&0\\
\eta_g&\eta_g^*&1
\end{array}\right)
\end{equation}
Similarly, the middle term $(PQ)$ represents any initial state and any final state so it will be a matrix $(pq)'$ with all elements equal to $\kappa$. This is the general propagator for a single MUB step. Transforming it to $(pq)$ we have:
\begin{equation}\label{eq:pqchange}
(pq) = \kappa \left(\begin{array}{ccc}
1&\eta_g&\eta_g^*\\
\eta_g^*&1&\eta_g\\
\eta_g&\eta_g^*&1
\end{array}\right).
\end{equation}
Finally, $(QX)$ will be $\kappa$ times a matrix whose left ($x$) column only is nonzero, with all entries 1. It converts to:
\begin{equation}
(qx) = \kappa \left(\begin{array}{ccc}
1&0&0\\
\eta_g^*&0&0\\
\eta_g&0&0
\end{array}\right),
\end{equation}
and the sum over paths is represented by the matrix product:
\begin{equation}\label{eq:G3ZXdone}
G_2(+\vec{x},+\vec{z}) = \kappa^3 \left(\begin{array}{ccc}
0&0&0\\
0&0&0\\
\eta_g&\eta_g^*&1
\end{array}\right)
\left(\begin{array}{ccc}
1&\eta_g&\eta_g^*\\
\eta_g^*&1&\eta_g\\
\eta_g&\eta_g^*&1
\end{array}\right)\left(\begin{array}{ccc}
1&0&0\\
\eta_g^*&0&0\\
\eta_g&0&0
\end{array}\right).
\end{equation}
Upon multiplying, and factoring out the $ZX$ matrix to the right, we obtain:
\begin{equation}
\begin{array}{rcl}
G_2(+\vec{x},+\vec{z}) &=& 
3\kappa^3[(\eta_g+\eta_g^*\eta_g^*+\eta_g^*\eta_g^2)/\eta_g]
\left(\begin{array}{ccc}
0&0&0\\
0&0&0\\
\eta_g&0&0
\end{array}\right).
\end{array}
\end{equation}
Replacing $\eta_g$ with $\sqrt{1/2}\exp(i\pi/12)w^g$, we find that
\begin{equation}
3\kappa^3(\eta_g+\eta_g^*\eta_g^*+\eta_g^*\eta_g^2)/\eta_g = \kappa^3(6 - 3i/2).
\end{equation}
The result does not depend on $g$ and is the same as the sum over paths, Eq.~(\ref{eq:SumOverPathsExample}).

The above result generalizes. Longer path integrals introduce extra factors of the $(pq)$ array of Eq.~(\ref{eq:pqchange}) and extra factors of $\kappa$. Define:
\begin{equation}
G_g = \kappa
\left(\begin{array}{ccc}
1&\eta_g&\eta_g^*\\
\eta_g^*&1&\eta_g\\
\eta_g&\eta_g^*&1
\end{array}\right),
\end{equation}
so longer paths will involve powers of $G_g$. Note that $G_g$ is 1-circulant, that is each row is identical to the row above, but rotated 1 position to the right.

The discrete Fourier transform diagonalizes 1-circulant matrices. This allows us to compute $G_g^N$ by taking the discrete Fourier transform, taking powers of the diagonal entries, and then reverse transforming. Thus we can solve for $G_g^N$ in closed form.

Define the Fourier transform matrix $F$ as
\begin{equation}\label{eq:FT}
F = \frac{1}{\sqrt{3}}\left(\begin{array}{ccc}
w&w^*&1\\
w^*&w&1\\
1&1&1
\end{array}\right),
\end{equation}
so that the discrete Fourier transform of a vector $\vec{v}$ is $F\vec{v}$. Then the discrete Fourier transform of a matrix $M$ is
\begin{equation}
\tilde{M} = F\;M\;F^*.
\end{equation}
Given a 1-circulant matrix with top row $(A,B,C)$, the discrete Fourier transform converts it to a diagonal matrix:
\begin{equation}
\left(\begin{array}{ccc}
A+w^*B+wC&0&0\\
0&A+wB+w^*C&0\\
0&\;0\;&A+B+C
\end{array}\right).
\end{equation}
The elements down the diagonal are $\sqrt{3}$ times the discrete Fourier transform of the vector $(C,B,A)^t$. The transform of $G_g$ is a diagonal matrix $\tilde{G}_g$ of this form. The $j$th element on the diagonal of $\tilde{G}_g$ is:
\begin{equation}\label{eq:GgA}
[\tilde{G}_g]_{jj} = \kappa [1+\sqrt{2}\cos(2g\pi/3 - 2j\pi/3 + \pi/12)].
\end{equation}
The three matrices $\tilde{G}_g$ have the same diagonal elements but in an order that depends on $g$.

The largest entry on the diagonal of $\tilde{G}_g$ is $[\tilde{G}_g]_{gg}$; it will dominate $(\tilde{G}_g)^N$ as $N\to \infty$. In order for the limit to exist, this diagonal entry must be $1$. Therefore we have that 
\begin{equation}
1 = [\tilde{G}_g]_{gg} = \kappa[1+\sqrt{2}\cos(\pi/12)].
\end{equation}
The other two diagonal entries in $\tilde{G}_g$ are $[\tilde{G}_g]_{g+1,g+1} =$  $\kappa(1+\sqrt{2}\cos(-7\pi/12))$ $= 2-\sqrt{3}$ and $[\tilde{G}_g]_{g+2,g+2} =$ $\kappa[1+\sqrt{2}\cos(-15\pi/12)] = 0$. We have:
\begin{equation}
G_1^N = F^*\;\left(\begin{array}{ccc}
1&0&0\\
0&(2-\sqrt{3})^N&0\\
0&0&0\end{array}\right)\;F,
\end{equation}
and similarly for $G_2^N$ and $G_3^N$, but with the elements on the diagonal rotated. Finally, to obtain the non commutative amplitudes, one performs the reverse transformation.

In the limit as $N$ goes to infinity, $\tilde{G}_g^N$ becomes a diagonal matrix with the $g$th entry equal to one and all other entries zero. Taking the inverse discrete Fourier transform we find:
\begin{equation}\label{eq:Ginfinity}
G_g^{\infty} = F^* (\tilde{G}_g)^{\infty} F = 
\frac{1}{3}\left(\begin{array}{ccc}
1&w^{-g}&w^{+g}\\
w^{+g}&1&w^{-g}\\
w^{-g}&w^{+g}&1
\end{array}\right).
\end{equation}
The magnitude of all the entries are $1/3$ so the transition probabilities all equal $(1/3)^2 = 1/9$. On converting the above to non commutative form, we find:
\begin{equation}\label{eq:Gginfinity}
G_g^{'\infty} = \frac{1}{3\sqrt{2}}\left(\begin{array}{ccc}
\sqrt{2}&e^{-i\pi/12}w^{-g}&e^{+i\pi/12}w^{+g}\\
e^{+i\pi/12}w^{+g}&\sqrt{2}&e^{-i\pi/12}w^{-g}\\
e^{-i\pi/12}w^{-g}&e^{+i\pi/12}w^{+g}&\sqrt{2}
\end{array}\right).
\end{equation}
The above can be translated from the matrix using the path basis to obtain:
\begin{equation}\label{eq:Ggpathbasis}
\begin{array}{rcl}
G_g^{'\infty} &=& [(X+Y+Z) + e^{-i\pi/12}w^{-g}\sqrt{1/2}(XY+YZ+ZX)\\
&& + e^{+i\pi/12}w^{+g}\sqrt{1/2}(YX+ZY+XZ)]/3.
\end{array}
\end{equation}
We will associate these three propagators with the three generations of elementary fermions.


\section{The Particle Generations}\label{sec:Generations}

The long-time propagators $G_g^{'\infty}$ of Eq.~(\ref{eq:Ggpathbasis}) are idempotent:
\begin{equation}
(G_g^{'\infty})^2 = G_g^{'\infty},
\end{equation}
they annihilate each other:
\begin{equation}
G_g^{'\infty}G_h^{'\infty} = 0,\;\;\textrm{if $g\neq h$,}
\end{equation}
and have unit traces. The off diagonal elements of $G_g^{'\infty}$ are each multiplied by a factor of $w^{+g}$ or $w^{-g}$. Since the sum of powers of $w$ is zero, i.e,
\begin{equation}
w+w^2+w^3 = w^{-1} + w^{-2} + w^{-3} = 0,
\end{equation}
the three matrices $G_g^{'\infty}$ sum to the unit matrix:
\begin{equation}\label{eq:SumTo1}
G_1^{'\infty}+G_2^{'\infty}+G_3^{'\infty} = X + Y + Z = 1.
\end{equation}
These are the same relationship Eq.~(\ref{eq:Idempotency}) to Eq.~(\ref{eq:TraceIs1}), exhibited by the projection operators for a complete set of basis states. Therefore, ignoring the short-time behavior, the long-time propagators define a 3-dimensional Hilbert space. The three projection operators $\bar{X}$, $\bar{Y}$, and $\bar{Z}$ define another 3-dimensional Hilbert space. This is just three times the number of states in the usual spin-1/2. We associate the tripling with the three generations.

The factors $w^{+g}$ and $w^{-g}$ that distinguish the three generations are cubed roots of unity. Since the long-time propagators are otherwise identical, this suggests that the generations should be simple when characterized as cube roots of unity. That is, $w^{+g}$ are the three solutions to the equation:
\begin{equation}
z^3 = 1.
\end{equation}
Better, the off diagonal elements of Eq.~(\ref{eq:Gginfinity}) are
\begin{equation}\label{eq:ComplexRootsU}
u_g = \exp(2ig\pi/3 + i\pi/12)/\sqrt{2}.
\end{equation}
These are the roots of the equation $z^3 = (1+i)/4$. We expect that the differences between the generations of elementary particles should be simple when expressed as functions of $u_g$.

What we have done amounts to complexifying generation; we extend generation from a discrete variable that takes on the three values 1, 2, and 3, to a complex variable that can take on any complex value. This extension of generation amounts to consideration of particle states that are not the orthogonal long-time propagators. This is similar to the work of T. Regge who extended orbital angular momentum to a complex variable. \cite{Regge1959}


Our long time propagators contain extra information beyond what is present in a spin-1/2 state. Instead of a spin axis, we are provided instead with a set of three orthogonal unit vectors $\vec{x}$, $\vec{y}$, and $\vec{z}$. From these we obtain the spin axis as the normalized sum:
\begin{equation}
\vec{u} = (\vec{x}+\vec{y}+\vec{z})/\sqrt{3}.
\end{equation}
The projection operator for spin in this direction $\vec{u} = (1,1,1)/\sqrt{3}$ is
\begin{equation}\label{eq:projspinu}
\sigma_u = (1+(\sigma_x+\sigma_y+\sigma_z)/\sqrt{3})/2.
\end{equation}
Our problem is that other complete sets of MUBs will give the same spin axis and spin projection operator.

Given two spin-1/2 states $|+u\ra$ and $|+v\ra$, the transition probability between them is given by the square of the inner product:
\begin{equation}\label{eq:uvuspinhalf}
|\la+u|+v\ra|^2 = \textrm{tr}(|+u\ra\la+u| |+v\ra\la+v| |+u\ra\la+u|).
\end{equation}
The right hand side of the above is written in terms of matrix multiplication of pure density matrices. Our long time propagators Eq.~(\ref{eq:Ginfinity}) are written in matrix form and their concatenation is given by matrix multiplication so we can use this form to verify that the long time propagators lead to the usual spin-1/2 relations.

In computing Eq.~(\ref{eq:uvuspinhalf}), we need to make sure that the rows and columns of our matrices are compatible in terms of their orientations. In general, $|+u\ra$ and $|+v\ra$ will not share the same complete MUB sets so we cannot repeat our trick, used in Eq.~(\ref{eq:MakeLegal}), of arranging for the matrix multiplication to be between identical bases. Instead, we will have to sum over orientations.

MUBs must remain perpendicular; consequently there are only three orientations possible -- those corresponding to the even permutations of the three bases:
\begin{equation}\label{eq:evenperms}
\{xyz, yzx, zxy\}.
\end{equation}
The nine complex numbers in the long time propagators are left unchanged by these permutations. Consequently, when we sum over these permutations, the result will be alterations of the path basis Eq.~(\ref{eq:PathBasis}). For example, the three orientations will take the path basis element $X$ to $\{X,Y,Z\}$, so the result of the sum will be $X\to X+Y+Z$, which can be rewritten as $(3+\sigma_x+\sigma_y+\sigma_z)/2$. Similarly, $Y$ and $Z$ will be taken to $X+Y+Z$. The complete transformation on the path basis is:
\begin{equation}\label{eq:summedbasis}
\begin{array}{ccccl}
X,Y,Z &\to& X+Y+Z &=& (3+(\sigma_x+\sigma_y+\sigma_z))/2,\\
XY,YZ,ZX &\to& XY+YZ+ZX &=& (3+(2+i)(\sigma_x+\sigma_y+\sigma_z))/4,\\
YX,ZY,XZ &\to& YX+ZY+XZ &=& (3+(2-i)(\sigma_x+\sigma_y+\sigma_z))/4.
\end{array}
\end{equation}
The right hand sides of the above are linear combinations of 1 and $(\sigma_x+\sigma_y+\sigma_z)$ which is close to what we need, Eq.~(\ref{eq:projspinu}).

Summing over orientations will reduce the $3\times 3$ matrix form of a long time propagator to a $1\times 1$ form, that is, it will be reduced to an object that can be represented among the usual complex $2\times 2$ matrices. The result Eq.~(\ref{eq:summedbasis}) shows that this object will be some complex linear combination of 1 and $\sigma_u$.

Before we can compute the spin-1/2 interaction between two long time propagators we need to renormalize the long time propagators for spin-1/2 type interactions with themselves. That is, we need to make sure that we have the long time propagators in a form where they will satisfy the projection operator rule $G_aG_a = G_a$ where $G_a$ is mixed over orientation. Given that $G_a$ will be a linear combination of 1 and $\sigma_u$, there are only four possible solutions to this equation
\begin{equation}\label{eq:posssolns}
0,\;\;(1\pm\sigma_u)/2,\;\;1.
\end{equation}
These can be found by solving the equation $(\alpha+\beta\sigma_u)^2 = \alpha+\beta\sigma_u$ for complex constants $\alpha$ and $\beta$.

To see which of the solutions Eq.~(\ref{eq:posssolns}) is the one obtained when the long time propagator $G_a = \alpha+\beta\sigma_u$ interacts with itself we look at the long time limit. Noting that $\sigma_u^2=1$, we find that
\begin{equation}
\lim_{n\to\infty} (G_a)^n = \lim_{n\to \infty}
\frac{(\alpha+\beta)^n + (\alpha-\beta)^n}{2}+
\frac{(\alpha+\beta)^n - (\alpha-\beta)^n}{2}\sigma_u.
\end{equation}
To make this limit finite and nonzero we renormalize $G_a$ by dividing it by $\alpha+\beta$ or $\alpha-\beta$, whichever has the larger magnitude. The result will have a limit of 0 or 1 only when $\beta=0$, a set of measure zero (which does not include our propagators). Otherwise the limit will be $(1\pm\sigma_u)/2$. From this the spin-1/2 relations given in Eq.~(\ref{eq:uvuspinhalf}) follow.

\section{Lepton Masses}\label{sec:LepMass}

In the standard model, mass is an interaction between the left and right handed spin-1/2 states. The mass terms in the Lagrangian, for the three generations, are:
\begin{equation}\label{eq:StandLagrang}
\Sigma_{g=1}^3 (m_g\psi_{gL}^*\psi_{gR} + m_g\psi_{gR}^*\psi_{gL}).
\end{equation}
where $m_g$ is the mass of the $g$th generation particle. When the $\psi_{gL}$ and $\psi_{gR}$ are rewritten in terms of this paper's long-time propagators, it's natural to expect that generational differences in mass will be due to differences in the propagators. This provides hope that the three experimentally determined constants, $m_g$ can be united into a single constant. The differences between $m_g$ will then be determined by differences in the corresponding long-time spin propagators.

The various portions of the long-time propagators differ only in that they depend on $w^g$. Consequently, it may be useful to write $m_g$ in the form:
\begin{equation}\label{eq:MassFormulag}
m_g = \Sigma_{n=1}^3A_nw^{ng}.
\end{equation}
Since the masses are real, we have that $A_1^* = A_2$ and $A_3$ is real. Putting $A_1 = B\exp(iC)$ we have:
\begin{equation}\label{eq:FMass}
\begin{array}{rcl}
m_g &=& A_3 + A_1\exp(2ig\pi/3) + A_2\exp(-2ig\pi/3),\\
&=& A_3 + 2B\cos(2g\pi/3 + C).
\end{array}
\end{equation}
In units with $c=1$, this is the charged lepton mass equation used by Gerald Rosen:
\begin{equation}\label{eq:GRosenEqn}
\sqrt{m_g} = 17716\;\sqrt{\textrm{eV}}\;[1+\sqrt{2}\cos(2g\pi/3 + 2/9)],
\end{equation}
who notes that it is accurate to $O(10^{-5})$ and relates it to a Dirac-Goldhaber model of the quarks and leptons. \cite{GRosen} In this formula, $C=2/9$ and $B=A_3\sqrt{1/2}$. The above is similar to the long-time propagators in that it includes a square root of 2, but different in that the angle $\pi/12$ is replaced by $2/9$.

In 1982 \cite{Koide82,Koide83}, Yoshio Koide discovered a formula for the charged lepton masses:
\begin{equation}
2(\sqrt{m_{e}} + \sqrt{m_{\mu}} + \sqrt{m_{\tau}})^{2} =
3(m_{e} + m_{\mu} + m_{\tau}).
\end{equation}
Since then, measurements of the $\tau$ mass have converged to Koide's prediction; it is still accurate to well within experimental error. This formula follows from the $\sqrt{2}$ in Eq.~(\ref{eq:GRosenEqn}).

The neutrino masses are known by the differences in the squares of their masses implied by neutrino oscillation measurements. There are only two such measurements, ``solar'' and ``atmospheric'', so we need another restriction to predict the neutrino masses. Koide's formula would provide the third restriction but it is incompatible with the oscillation measurements. On the other hand, the oscillation measurements \emph{are} compatible with Eq.~(\ref{eq:FMass}) with $A_3/B = \sqrt{2}$. The resulting equation for the neutrino masses
\begin{equation}\label{eq:BrannenEqn}
\sqrt{m_{\nu g}} = 0.1000(26)\sqrt{\textrm{eV}}\;[1+\sqrt{2}\cos(2g\pi/3 +\pi/12 + 2/9].
\end{equation}
gave predictions for the neutrino masses in 2006: \cite{MASSES2}
\begin{equation}\label{eq:BrannPredicts}
\begin{array}{rcl}
m_{\nu 1} &=& 0.00038,\\
m_{\nu 2} &=& 0.0089,\\
m_{\nu 3} &=& 0.0507,
\end{array}
\end{equation}
which give differences in the squares of the masses as:
\begin{equation}
\begin{array}{rcrcl}
\nabla m^2_{\textrm{sol}} &=& |m_{\nu 1}^2 - m_{\nu 2}^2| &=& 7.9 \times 10^{-5}\;\textrm{eV}^2,\\
\nabla m^2_{\textrm{atm}} &=& |m_{\nu 2}^2 - m_{\nu 3}^2| &=& 2.5 \times 10^{-3}\;\textrm{eV}^2.
\end{array}
\end{equation}
The mass predictions satisfy Koide's equation but with $-\sqrt{m_{\nu 1}}$.

A recent measurement of the solar neutrino mass parameter $\nabla m^2_{\textrm{sol}}$ from the Sudbury Neutrino Observatory \cite{Sudbury2009} is $7.59(21)\times 10^{-5}\;\textrm{eV}^2$. Data from MINOS \cite{Minos2009} give $\nabla m^2_{\textrm{atm}}$ around $2.43\times 10^{-3}\;\textrm{eV}^2$, so the above mass predictions are still well within the error bars of the oscillation measurements. Adjusting the mass scale $0.1000(26)$ a little to match the new experimental data, a more current neutrino mass formula is:
\begin{equation}\label{eq:BrannenEqn2}
\sqrt{m_{\nu g}} = 0.0990\;\sqrt{\textrm{eV}}\;[1+\sqrt{2}\cos(2g\pi/3 +\pi/12 + 2/9)],
\end{equation}
which gives $ 7.7\times 10^{-5}$ and $2.44\times 10^{-3}\;\textrm{eV}^2$.

The two lepton mass formulas, Eq.~(\ref{eq:GRosenEqn}) and Eq.~(\ref{eq:BrannenEqn2}), are similar except for the mass scale and the angle. The two formulas have the common angle $2/9$, a number which is close to the Cabibbo angle. They differ in that the neutrino takes the angle $\pi/12$ that appears in the long-time projection operators Eq.~(\ref{eq:Gginfinity}). Perhaps the charged lepton mass interaction is a simple interaction while the neutrino mass interaction is more complicated. The $\pi/12$ arises due to quantum phase. It may be useful to note that massless spin-1 particles have the same quantum phase as spin-1/2 particles, while spin-0 particles have no corresponding quantum phase.

The obvious way to obtain the charged lepton masses in terms of the $u_g$ of Eq.~(\ref{eq:ComplexRootsU}) is:
\begin{equation}\label{eq:GrosenUsingU}
\sqrt{m_g} = 17716\;\sqrt{\textrm{eV}}\; [1 + u_ge^{2i/9-i\pi/12} + u_g^*e^{-2i/9+i\pi/12}],
\end{equation}
the corresponding equation for the neutrinos is:
\begin{equation}\label{eq:BrannenEqnUsingU}
\sqrt{m_{\nu g}} = 0.0990\;\sqrt{\textrm{eV}}\;[1 + u_ge^{2i/9} + u_g^*e^{-2i/9}],
\end{equation}
The numerical constants in the above two equations, $17716$ and $0.0990$, have a ratio very close to an exact power of three. That is, $17716/0.0990 = 3^{11.009}$. We speculate that a theory explaining the mass hierarchy between the charged and neutral leptons will involve a coupling constant that is a power of 3 (such as the 1/3 of Eq.~(\ref{eq:Ginfinity})), and that differences in the complexity of the mass interaction explains the small neutrino mass.

\section{Discussion}\label{sec:Discussion}

In ontological theories of quantum mechanics, one attempts to choose which mathematical treatment of quantum mechanics is a model of reality in that the elements of the model correspond directly to elements of Nature. If we choose spin and position as ontological observables, then it is natural for us to desire that they have similar relationships between their classical and quantum behavior.

The calculation is robust in that the long-time propagators $G_g^{\infty}$ of Eq.~(\ref{eq:Gginfinity}) do not depend on the details of $G_g$. For example, one could consider a theory where the paths require all adjacent propagators to be different. This would allow paths like $XYZX$ but disallow $XXYZ$. This change is accomplished by putting zeros on the diagonal of $G_g$. The result would be that $G_g^N$ and $\kappa$ would be different, but the discrete Fourier transform of $G_g$ would still be diagonal with one element larger in magnitude than the others and this element would dominate $G_g^N$ leaving $G_g^{\infty}$ unchanged. Spin-1/2 behavior is similarly robust.

The calculation uses the spin-up projection operators $\{X,Y,Z\}$, and treats them equally but does not include the spin-down projection operators $\{\bar{X},\bar{Y},\bar{Z}\}$. Summing over all paths between orthogonal observables such as $Z$ and $\bar{Z}$ must give zero by symmetry considerations. For example, $ZX\bar{Z} + Z\bar{X}\bar{Z}$ $=Z(X+\bar{X})\bar{Z}$ $=Z(1)\bar{Z} = 0$. Thus one cannot include sums over all paths arbitrarily in a spin path integral. More general long-time propagator solutions can be obtained by the usual methods of rotation and linear superposition.

MUBs are a fundamental object of interest in quantum information theory. They encapsulate the essence of the relationship between complementary observables such as position and momentum. To find MUBs at the foundation of position path integrals, spin-1/2, and the particle generations may be more surprising to theorists in elementary particles than to those in quantum information.

In modern elementary particle theory spin arises as a result of examining the irreducible representations of the homogeneous Lorentz group \cite{WeinbergI}, that is, spin-1/2 is one of the few possibilities allowed in the intersection of the special theory of relativity with quantum mechanics. Physics has found this intersection a bountiful place to look for elementary particle models. Despite these successes, there has been difficulty combining gravitation with quantum mechanics.

Our best theories, when extrapolated to very small distances, predict that space is anything but flat. Lorentz invariance cannot possibly apply at short distances so it cannot logically be used to restrict theories in that regime. Instead, the assumptions of this paper must be judged on the basis of how arbitrary they are, and whether the resulting calculations are compatible with observations.

Situations involving extremely non flat spacetime are available, at least theoretically, in that we can look at the dynamics of black holes. Non rotating black holes exponentially decay to spherical symmetry. Presumably, their exponential approach is accomplished by the radiation of elementary particles. Accordingly, L. Motl \cite{MotlTPS} examined the vibration modes of black holes for various spin cases. In addition to the expected results for spin-1/2 and spin-1, he found a spectrum of vibrations he called ``tripled Pauli statistics''. These were the results of the spin-0 and spin-2 vibration modes. Perhaps they have something to do with the existence of three generations of elementary fermions.

\section{Acknowledgments}\label{sec:Ack}

The author thanks his parents for financial assistance and Marni Sheppeard for advice and encouragement, and the anonymous reviewers for improvements in the manuscript.

%
%


\bibliographystyle{unsrtnat}
\bibliography{spinpath}   

%
%

\end{document}